**Hybridization Gap and Edge States in Strained-Layer $InAs/In_{0.5}Ga_{0.5}Sb$ Quantum Spin Hall Insulator**

Wenfeng Zhang[1†], Peizhe Jia[1†], Wen-kai Lou[2], Xinghao Wang[1], Shaokui Su[3], Kai Chang[4], and Rui-Rui Du[1,5]*

[1]International Center for Quantum Materials, School of Physics, Peking University, Beijing 100871, China

[2]State Key Laboratory of Semiconductor Physics and Chip Technologies, Institute of Semiconductors, Chinese Academy of Sciences, Beijing 100083, China

[3]Institute of Physics, Chinese Academy of Sciences, Beijing 100190, China.

[4]Center for Quantum Matter, Zhejiang University, Hangzhou 310027, China

[5]Hefei National Laboratory; Hefei 230088, China

† Both authors contribute equally to this work.

* Corresponding author. Email: rrd@pku.edu.cn

The hybridization gap in strained-layer $InAs/In_xGa_{1-x}Sb$ quantum spin Hall insulators (QSHIs) is significantly enhanced compared to binary InAs/GaSb QSHI structures, where the typical indium composition, x, ranges between 0.2 and 0.4. This enhancement prompts a critical question: to what extent can quantum wells (QWs) be strained while still preserving the fundamental QSHI phase? In this study, we demonstrate the controlled molecular beam epitaxial (MBE) growth of highly strained-layer QWs with an indium composition of x = 0.5. These structures possess a substantial compressive strain within the $In_{0.5}Ga_{0.5}Sb$ QW. Detailed crystal structure analyses confirm the exceptional quality of the resulting epitaxial films, indicating coherent lattice structures and the absence of visible dislocations. Transport measurements further reveal that the QSHI phase in $InAs/In_{0.5}Ga_{0.5}Sb$ QWs is robust and protected by time-reversal symmetry. Notably, the edge states in these systems exhibit giant magnetoresistance when subjected to a modest perpendicular magnetic field. This behavior is in agreement with the $Z_2$ topological property predicted by the Bernevig-Hughes-Zhang (BHZ) model, confirming the preservation of topologically protected edge transport in the presence of enhanced bulk strain.

---

## I. INTRODUCTION

Quantum spin Hall insulators (QSHIs), also referred to as two-dimensional topological insulators protected by time-reversal symmetry (TRS), are a topic of active research[1-4]. QSHIs feature an insulating bulk state and TRS-protected helical edge states at the sample perimeter. Among QSHI systems, inverted InAs/GaSb quantum wells (QWs) are notable for their tunable bulk and edge properties, which can be controlled via quantum well thickness and electric fields[5-11]. Nevertheless, the original InAs/GaSb QWs present two key limitations that affect both fundamental research and practical applications: the hybridization gap is relatively small, typically around 4 meV (about 50 K), and certain residual states within the hybridization gap compromise bulk insulating behavior[6, 7, 12-14]. These material constraints restrict the realization of robust QSH states in InAs/GaSb QWs.

Strained-layer $InAs/In_xGa_{1-x}Sb$ quantum wells exhibit hybridization gaps that are significantly larger than those found in conventional, unstrained InAs/GaSb heterostructures. The introduction of strain engineering in these QWs enhances the gap energies by a factor of three to eight, resulting in values ranging from 150 K to 400 K compared to approximately 50 K in the unstrained counterparts.[15-18]. This substantial increase in the hybridization gap is a key advantage of strained layer designs, as it improves both the bulk insulating properties and the robustness of the QSH effect in these systems. The enhanced hybridization gap, coupled with high-quality interfaces and coherent lattice structures, enables the realization of quantum wells that are more suitable for fundamental research and practical applications involving topological insulators.

Given the advantages of strained layers, an important question arises: how far can strain be increased while still preserving the fundamental QSHI phase? Each 20% increment of indium corresponds to approximately 1.25% strain in $In_xGa_{1-x}Sb$ QW, with reported indium content ranging from 20% to 40%[18]. The stability of higher indium percentages and the resulting bulk and edge state properties require investigation. Here, we report the MBE growth of QWs with 50% indium—the highest percentage reported to date. We demonstrate that these QWs possess coherent lattice structure and interfaces without visible dislocations, and that both the gap energy and property of edge states align with theoretical predictions. Additionally, magnetoresistance for perpendicular magnetic fields shows significant enhancement, potentially indicating gap opening under broken TRS. For comparison, we also grew samples with 25% indium and studied their magnetoresistance. Previous studies on edge state transport have only reached up to 32% indium[19].

In this study, we have successfully fabricated strained-layer $InAs/In_xGa_{1-x}Sb$ quantum wells with indium compositions of x = 0.5 and 0.25 using MBE growth. In our experiments, structure characterizations employed reflection high energy electron diffraction（RHEED）, scanning transmission electron microscope (STEM), and atomic force microscope (AFM). Electrical transport measurements employed an AC lock-in technique (17Hz) in a 300 mK $^3$He refrigerator with 9 T superconductor magnet, a base temperature 20 mK dilution refrigerator with 18 T superconductor magnet, and 9 T physical property measurement system （PPMS）.

## II. MBE GROWTH AND STRUCTURAL CHARACTERIZATIONS

Wafers were grown in a RIBER C21DZ MBE system equipped with valved crackers for $Sb_2$ and $As_4$. Group III elements were evaporated using standard Knudsen cells. N-type GaAs (001) substrates were used, first degassed in the buffer chamber for 1.5 hours at 400°C, then transferred to the growth chamber. Substrates were heated to 640°C under $As_4$ flux for 10 minutes to remove native oxide. After oxide desorption, a 250 nm GaAs layer and a 5 nm AlAs layer were grown at 620°C to smooth the surface. The substrate temperature was then lowered to 530°C. Before buffer layer growth, a 30 nm AlSb nucleation layer was deposited, followed by an $Al_{0.7}Ga_{0.3}Sb$ buffer layer to accommodate the 7% lattice mismatch between the GaAs substrate and quantum wells. The temperature was then reduced to 500°C for the remainder of the growth. Before growing the active region, a short-period superlattice [10 × (2.5 nm AlSb + 2.5 nm GaSb)] was grown to improve morphology and transport properties. After the superlattice, the active region was grown.

The active region consisted of barrier layers, QWs, and a 3 nm GaSb cap layer. The barrier was designed as a digital-alloy structure, alternating AlSb and GaSb layers to achieve a nominal composition of $Al_{0.75}Ga_{0.25}Sb$. The top and bottom barriers were 50 nm and 100 nm thick, respectively. Two types of QW structures were fabricated, as shown in Figs. 1(a) and 1(b): Sample A with an 8 nm InAs/3.6 nm $In_{0.5}Ga_{0.5}Sb$ structure, and Sample B with an 8 nm InAs/5 nm $In_{0.25}Ga_{0.75}Sb$ structure.

AFM analysis revealed root-mean-square (RMS) roughness values of 0.562 nm and 0.691 nm for samples A and B, respectively, as shown in Figs. 1(b) and (c). Both values exceed the typical monolayer roughness. Furthermore, both samples exhibited mound-like features associated with spiral growth around threading dislocations with screw character, consistent with prior reports for samples grown on GaAs substrates[20]. As shown in Figs. 1(d) and (e), STEM lattice images confirmed the coherence of the crystalline structure at all heterostructure interfaces, with no observable linear defects or point dislocations. The digital-alloy $Al_{0.75}Ga_{0.25}Sb$ barriers showed alternating dark and bright stripes, corresponding to the periodic AlSb and GaSb layers.

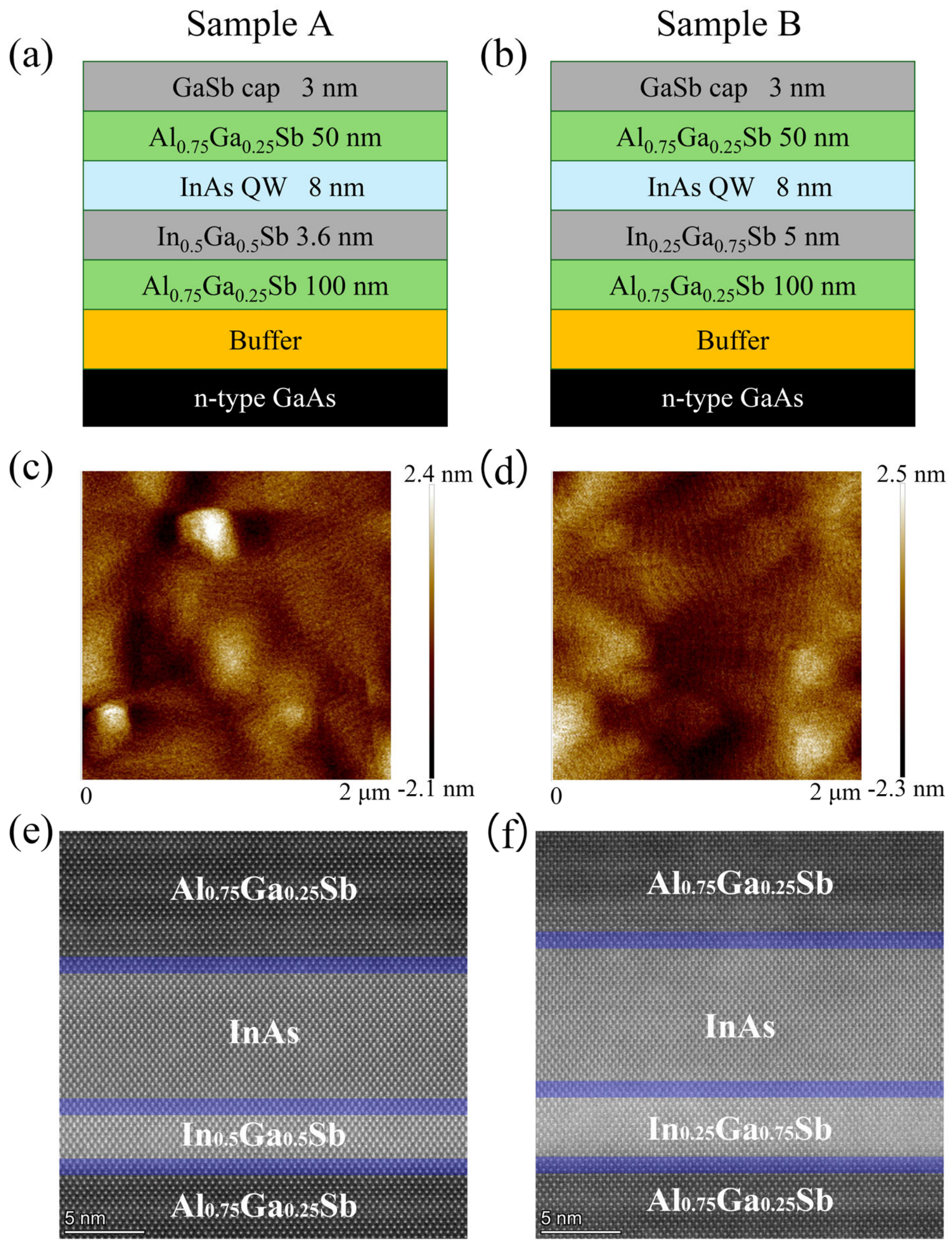

Fig. 1. (a) and (b) show the sample structure of the samples A and B, respectively. (c) and (d) show the representative AFM image of samples A and B, respectively. (e) and (f) show the cross-sectional STEM image of the active region of samples A and B, respectively. Blue line is a guide for the eyes.

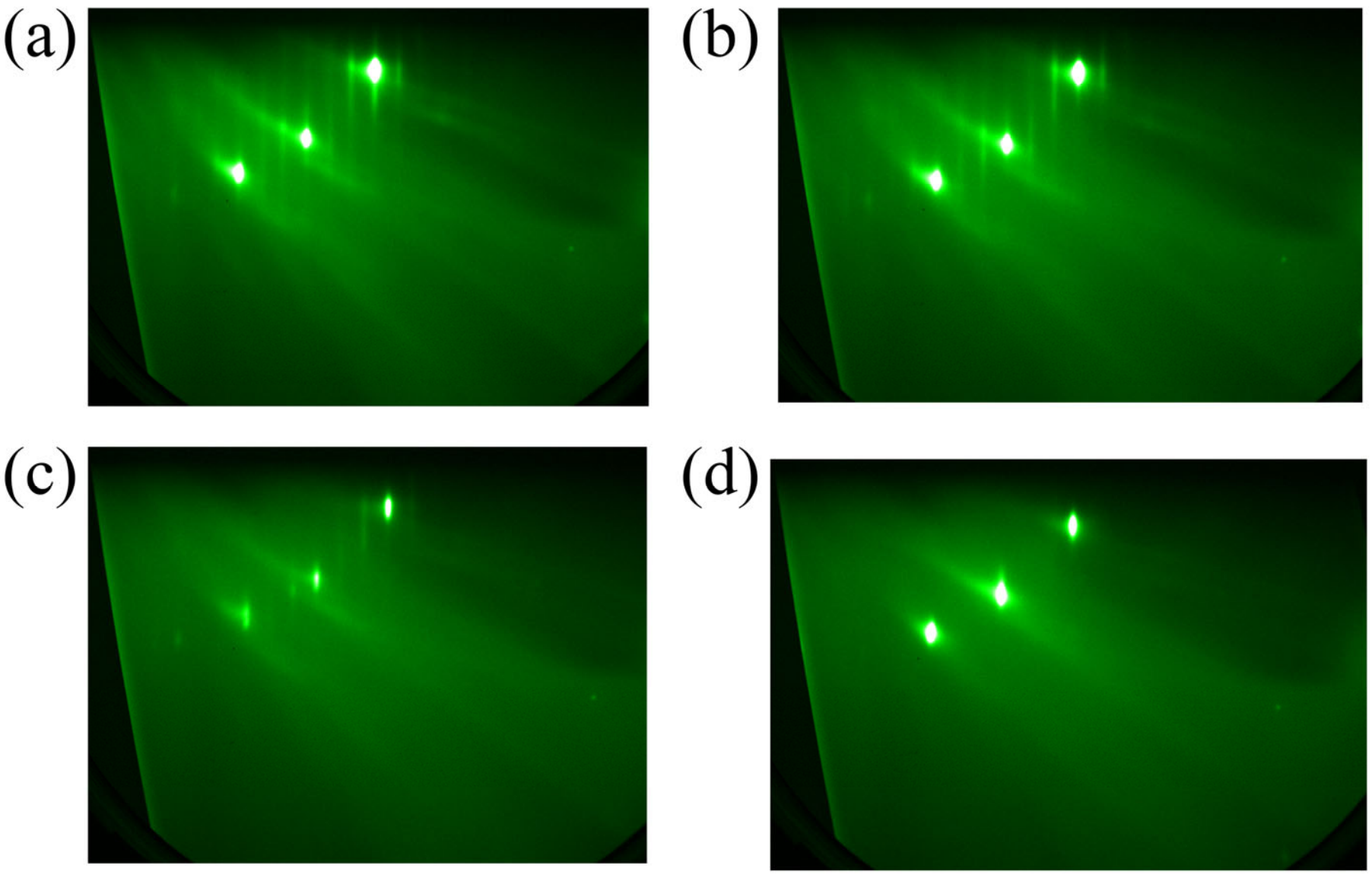

Fig. 2. RHEED patterns of representative epitaxial layer of $InAs/In_{0.5}Ga_{0.5}Sb$ QWs during the MBE growth. (a) and (b) show the AlSb layer and GaSb layer of the digital-alloy $Al_{0.75}Ga_{0.25}Sb$, respectively. (c) RHEED pattern of the $In_{0.5}Ga_{0.5}Sb$ layer. (d) RHEED pattern of InAs layer.

We have used the RHEED to study *in situ* crystalline quality and surface morphology $In_{0.5}Ga_{0.5}Sb$ QWs of MBE growth. Fig. 2 shows the RHEED patterns of representative epitaxial layer taken along the [-110] azimuth offset for a clearer view for reconstruction streaks of $InAs/In_{0.5}Ga_{0.5}Sb$ QWs during the MBE growth. Figs. 2(a) and 2(b) show the AlSb layer and GaSb layer of the digital-alloy $Al_{0.75}Ga_{0.25}Sb$, respectively. For the digital-alloy $Al_{0.75}Ga_{0.25}Sb$ barrier, the RHEED patterns of AlSb layer and GaSb layer were distinctive streaky with surface reconstruction of (1×3), which is similar to those of the conventional grown AlSb and GaSb. Fig. 2(c) displays the RHEED pattern of the $In_{0.5}Ga_{0.5}Sb$ layer, where the (1×3) surface reconstruction showed a weaker intensity. This attenuation can be attributed to lattice relaxation effects induced by the substantial lattice mismatch. Fig. 2(d) displays the RHEED pattern of InAs layer, showing streaky features characteristic of a (1×1) surface reconstruction. The RHEED analysis confirms a sustained two-dimensional layer-by-layer growth mode throughout deposition, with the absence of three-dimensional island formation. This growth behavior ensures good crystalline quality and atomically smooth surface morphology of strained-layer $InAs/In_{0.5}Ga_{0.5}Sb$ QW.

## III. GAP ENERGY OF THE BULK STATE

For this study, Hall bar and Corbino devices were fabricated using lithography and wet etching. Ohmic contacts were made by soldering pure indium at 270°C for Hall bars, or by evaporating Ti/Au followed by annealing at 275°C for 10 minutes for Corbino devices. A 100 nm thick $HfO_2$ layer was deposited via atomic layer deposition, serving as both the dielectric layer for the front gate and a passivation layer for the entire device. A 10 nm/90 nm Ti/Au front gate was then added, with the n-type GaAs substrate functioning as a back gate.

Band structure calculations [figs. 3(a) and (b)] using the eight-band Kane model for strained-layer $InAs/In_xGa_{1-x}Sb$ quantum wells indicate that a hybridization energy gap (Eg) of approximately 25.48 meV can be attained in 8 nm InAs/3.6 nm $In_{0.5}Ga_{0.5}Sb$ QWs, confirming a substantial hybridization-induced band gap. Dual-gated Corbino devices [inset in fig. 3(c)] were used to study the bulk properties, where the edge conductance is shunted, ensuring measured signals originate solely from bulk transport. As shown in fig. 3(c), bulk conductance measurements of sample A show conductivity minimum at the charge neutrality point (CNP), indicating the emergence of an energy gap and gate-tunable control of the CNP. At 300 mK, the bulk conductivity is negligible for back gate voltage $V_{\text{back}}$ = 8 V, and becomes somewhat higher (on the order of 10 μS per square) for $V_{\text{back}}$ = 0 V. These results confirm that strained-layer $InAs/In_{0.5}Ga_{0.5}Sb$ QWs display an insulating hybridization gap.

Fig. 3(d) shows the temperature-dependent conductance of the sample A at $V_{\text{back}}$ = 0 V. Arrhenius analysis at higher temperatures yield a gap value of approximately 244 K (about 21 meV) for $InAs/In_{0.5}Ga_{0.5}Sb$ QWs. The experimentally measured gap here is about 17.6% smaller than the calculated value. This difference can be attributed to the fact that the experimental values depend on the gate bias (which changes the degree of band inversion), which has not been taken into account in calculations.

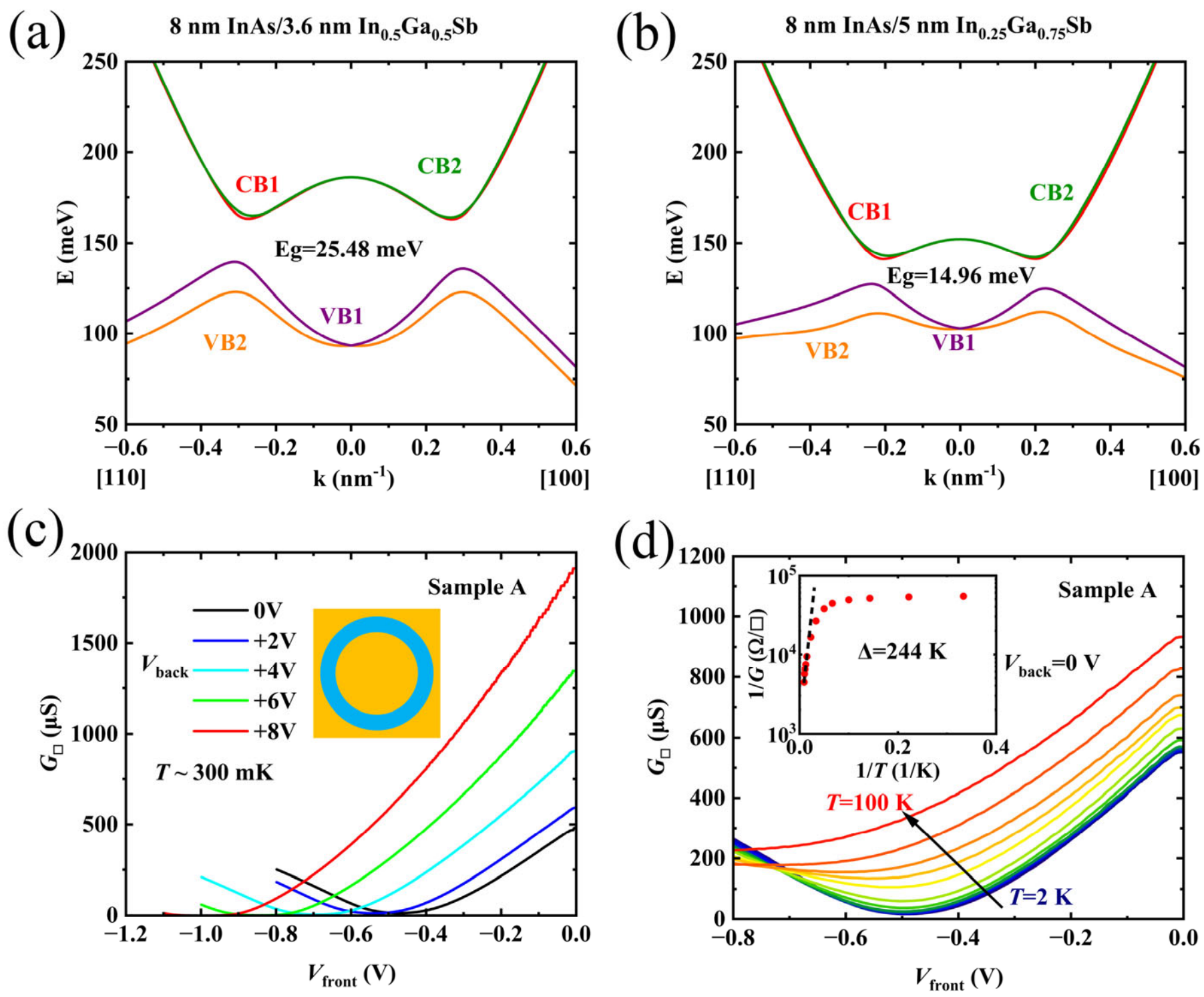

Fig. 3. Bulk transport of the strained-layer InAs/$In_{1-x}Ga_xSb$. (a) and (b) show the calculated bulk band structure of the InAs/$In_{0.5}Ga_{0.5}Sb$ QWs and InAs/$In_{0.25}Ga_{0.75}Sb$ QWs, respectively. (c) $G_\square$-$V_{front}$ traces of sample A measured in a Corbino device with various $V_{back}$, Corbino device configuration shown in inset. (d) The temperature-dependent conductance traces of sample A measured in a Corbino device at $V_{back} = 0$ V. The inset in (c) shows Arrhenius plots for sample A, energy gaps are deduced by fitting $G_{xx} \propto \exp\left(-\Delta/2k_BT\right)$, as shown by straight dash lines in the plot.

## IV. ELECTRICAL TRANSPORT OF EDGE STATES

The electrical quality of a typical InAs/$In_{0.5}Ga_{0.5}Sb$ QWs is characterized via Hall bar devices, and dual-gated Hall bar device shown in insert of Fig. 4(a). Fig. 4(a) shows the standard magnetotransport measurement of the sample A. Typical mobility for electrons is $2.4 \times 10^4$ $cm^2$/Vs at a density of $5 \times 10^{11}$ $cm^{-2}$. Well defined integer quantum Hall states and quantized Hall resistance are evident, which indicates the high quality of sample.

We have used dual-gated Hall bar device to study the properties of the edge state of the strained-layer $InAs/In_xGa_{1-x}Sb$ quantum wells. Fig. 4(b) shows the $R_{xx}$- $V_{front}$ traces of a 180 × 60 μm Hall bar device made by the sample A with various $V_{back}$ at $T \sim 300$ mK. As shown in Fig. 4(b), the $R_{xx}$ peak shifts systematically to a more negative $V_{front}$ with increasing $V_{back}$, also demonstrating gate-tunable control of the CNP. Moreover, the resistance peak values gradually increase with increasing $V_{back}$. Figs. 4(c) and 4(d) show the traces of longitudinal resistance $R_{xx}$ vs $V_{front}$ of a 40 × 20 μm Hall bar device made from the sample A, and sample B, with various perpendicular magnetic fields $B_{\perp}$ at $T \sim 20$ mK, respectively. Following the method described in Ref. [8, 21], we can estimate the edge state coherent length λ at the gate bias $V_{back} = 0$ V and $B_{\perp} = 0$ T for the samples A and B. The CNP $R_{xx}$ of Sample A is approximately 150 kΩ. With an edge length of 50 μm, this gives a ratio of ~3 kΩ/μm, corresponding to λ ~ 4.3 μm. For Sample B, we obtained λ ~ 8.1 μm. These data attested to the high quality of the sample.

Remarkably, the helical edge conductance of two samples both show strong magnetic field dependence. As shown in Figs. 4(c) and 4(d), the $R_{xx}$ peak resistance increases with perpendicular magnetic fields $B_{\perp}$, agreeing with magneto transport result in previous studies[15, 19]. In general, in the presence of an external magnetic field, TRS is broken, resulting in the opening of an energy gap in the helical edge states. Furthermore, additional scattering processes could emerge in the helical edge channel, such as the backscattering processes, and indeed, we observe that the helical edge conductance decreases under magnetic fields. These results provide evidence that the helical edge states are protected by TRS. We note that, remarkably, the $InAs/In_{0.5}Ga_{0.5}Sb$ QSHI edge state exhibits a giant magnetoresistance in response to a modest $B_{\perp}$, for instance, 10 fold increases under 0.6 T。

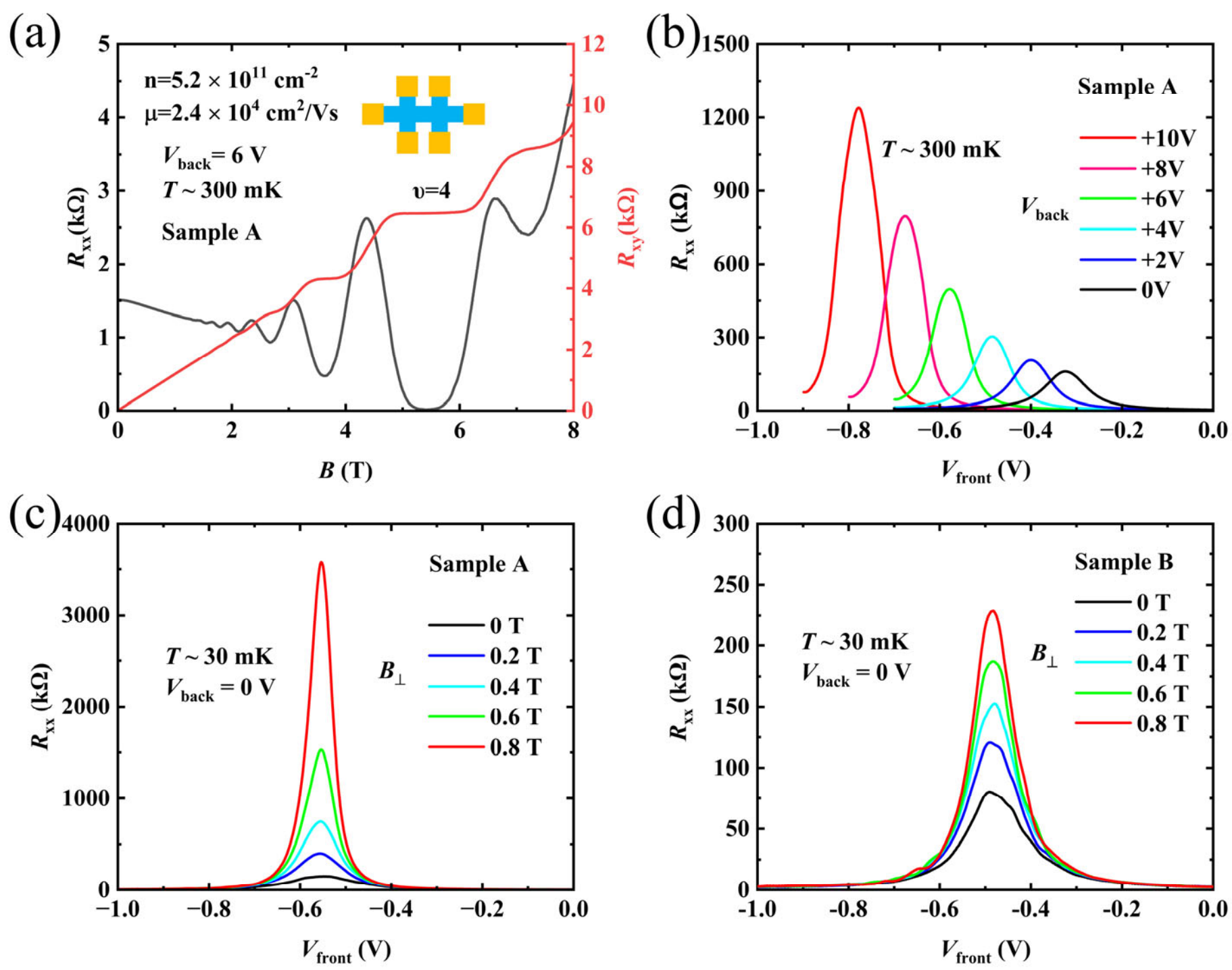

Fig. 4. Edge transport of the strained-layer InAs/$In_xGa_{1-x}Sb$. (a) Magnetoresistance and Hall resistance traces measured in sample A, Hall bar device configuration shown in inset. (b) $R_{xx}$-$V_{front}$ traces measured from Hall bar device made by the sample A with various $V_{back}$. (c) and (d) show the $R_{xx}$-$V_{front}$ traces of a Hall bar made by the sample A and sample B under different $B_{\perp}$, respectively.

## V. SUMMARY

Strained-layer InAs/$In_xGa_{1-x}Sb$ quantum wells with a higher indium composition of x = 0.5 were successfully fabricated using MBE growth. Structural characterization confirms excellent crystalline quality of the epitaxial films. Transport measurements establish that the QSHI phase in InAs/$In_xGa_{1-x}Sb$ quantum wells exhibit TRS protection. Remarkably, magnetoresistance of helical edge state between 0.2 T and 0.8 T reaches extraordinary values, opening experimental avenues for investigating TRS breaking and gap opening mechanisms in robust QSH systems, with potential applications in spintronics and topological devices.

## ACKNOWLEDGMENTS

We acknowledge initial experiments on molecular beam epitaxy of InAs/GaSb quantum wells at Peking University by Bingbing Tong and Siqi Yao, and thank Jianhua Zhao, Jianliang Huang, and Hong Lu for helpful discussions. Funding was provided by the Strategic Priority Research Program of Chinese Academy of Sciences (Grant No. XDB28000000 and No. XDB0460000), the Quantum Science and Technology-National Science and Technology Major Project (Grant No. 2021ZD0302600), and the National Key Research and Development Program of China (Grant No. 2024YFA1409002). A portion of this work was carried out at the Synergetic Extreme Condition User Facility (SECUF).

## REFERENCES

[1] Kane C L and Mele E J 2005 *Phys. Rev. Lett.* 95 226801

[2] Bernevig B A, Hughes T L and Zhang S-C 2006 *Science* 314 1757

[3] Hasan M Z and Kane C L 2010 *Rev. Mod. Phys.* 82 3045

[4] Qi X-L and Zhang S-C 2011 *Rev. Mod. Phys.* 83 1057

[5] Liu C, Hughes T L, Qi X-L, Wang K and Zhang S-C 2008 *Phys. Rev. Lett.* 100 236601

[6] Knez I, Du R R and Sullivan G 2010 *Phys. Rev. B* 81 201301

[7] Knez I, Du R-R and Sullivan G 2011 *Phys. Rev. Lett.* 107 136603

[8] Du L, Knez I, Sullivan G and Du R-R 2015 *Phys. Rev. Lett.* 114 096802

[9] Nichele F, Suominen H J, Kjaergaard M, Marcus C M, Sajadi E, Folk J A, Qu F, Beukman A J A, Vries F K d, Veen J v, Nadj-Perge S, Kouwenhoven L P, Nguyen B-M, Kiselev A A, Yi W, Sokolich M, Manfra M J, Spanton E M and Moler K A 2016 *New J. Phys.* 18 083005

[10] Jiang Y, Thapa S, Sanders G D, Stanton C J, Zhang Q, Kono J, Lou W K, Chang K, Hawkins S D, Klem J F, Pan W, Smirnov D and Jiang Z 2017 *Phys. Rev. B* 95 045116

[11] Shojaei B, McFadden A P, Pendharkar M, Lee J S, Flatté M E and Palmstrøm C

J 2018 *Phys. Rev. Mater.* 2 064603

[12] Suzuki K, Harada Y, Onomitsu K and Muraki K 2015 *Phys. Rev. B* 91 245309

[13] Mueller S, Pal A N, Karalic M, Tschirky T, Charpentier C, Wegscheider W, Ensslin K and Ihn T 2015 *Phys. Rev. B* 92 081303

[14] Naveh Y and Laikhtman B 2001 *EPL* 55 545

[15] Du L, Li T, Lou W, Wu X, Liu X, Han Z, Zhang C, Sullivan G, Ikhlassi A, Chang K and Du R-R 2017 *Phys. Rev. Lett.* 119 056803

[16] Irie H, Akiho T, Couëdo F, Suzuki K, Onomitsu K and Muraki K 2020 *Phys. Rev. Mater.* 4 104201

[17] Avogadri C, Gebert S, Krishtopenko S S, Castillo I, Consejo C, Ruffenach S, Roblin C, Bray C, Krupko Y, Juillaguet S, Contreras S, Wolf A, Hartmann F, Höfling S, Boissier G, Rodriguez J B, Nanot S, Tournié E, Teppe F and Jouault B 2022 *Phys. Rev. Res.* 4 L042042

[18] Akiho T, Couëdo F, Irie H, Suzuki K, Onomitsu K and Muraki K 2016 *Appl. Phys. Lett.* 109 192105

[19] Li T, Wang P, Sullivan G, Lin X and Du R-R 2017 *Phys. Rev. B* 96 241406

[20] Shojaei B, McFadden A, Shabani J, Schultz B D and Palmstrøm C J 2015 *Appl. Phys. Lett.* 106 222101

[21] Spanton E M, Nowack K C, Du L, Sullivan G, Du R-R and Moler K A 2014 *Phys. Rev. Lett.* 113 026804